\documentclass[%
 reprint,
 amsmath,amssymb,
 aps,
pra,
]{revtex4-2}

\usepackage{graphicx}% Include figure files
\usepackage{dcolumn}% Align table columns on decimal point
\usepackage[table]{xcolor}
\usepackage[normalem]{ulem}

\usepackage{bm}% bold math
\begin{document}

\preprint{APS/123-QED}

\newcommand{\SSS}[1]{{\color{magenta}[SSS: #1]}}
\newcommand{\DB}[1]{{\color{blue}[DB #1]}}
\newcommand{\AG}[1]{{\color{red}[AG #1]}}
\newcommand{\AK}[1]{{\color{teal}[AA #1]}}
\newcommand{\RA}[1]{{\color[rgb]{0,0.4,0}[RA: #1]}}

\title{Second-order correlations in directed emissions in sodium atoms}% Force line breaks with \\

\author{Ara Tonoyan}\thanks{}
\altaffiliation{}
\affiliation{Institute for Physical Research, National Academy of Sciences of Armenia, 0204 Ashtarak, Armenia}

\author{Sushree Subhadarshinee Sahoo}
\affiliation{Johannes Gutenberg-Universit{\"a}t Mainz, 55128 Mainz, Germany}
\affiliation{Helmholtz-Institut Mainz, 55099 Mainz, Germany}
\affiliation{Department of Physics, Indian Institute of Technology, Kanpur, Kanpur 208016, India}
\mbox{}
\author{Anahit Gogyan}
\email{agogyan@gmail.com}
\affiliation{Institute for Physical Research, National Academy of Sciences of Armenia, 0204 Ashtarak, Armenia }
\mbox{}
\author{Oleg Tretiak}
\affiliation{Johannes Gutenberg-Universit{\"a}t Mainz, 55128 Mainz, Germany}
\affiliation{Helmholtz-Institut Mainz, 55099 Mainz, Germany}
\mbox{}
\author{Razmik Aramyan}
\affiliation{Johannes Gutenberg-Universit{\"a}t Mainz, 55128 Mainz, Germany}
\affiliation{Helmholtz-Institut Mainz, 55099 Mainz, Germany}
\mbox{}
\author{Alexander Akulshin}
\affiliation{Optical Sciences Centre, Swinburne University of Technology, Melbourne 3122, Australia}
\mbox{}
\author{Dmitry Budker}
\affiliation{Johannes Gutenberg-Universit{\"a}t Mainz, 55128 Mainz, Germany}
\affiliation{Helmholtz-Institut Mainz, 55099 Mainz, Germany}
\affiliation{GSI Helmholtzzentrum für Schwerionenforschung GmbH, 64291 Darmstadt, Germany}
\affiliation{Department of Physics, University of California, Berkeley, California 94720, USA}

\date{\today}% It is always \today, today,

\begin{abstract}
We report on measurements of second-order intensity correlations $g^{(2)}(\tau)$ of infrared emission under bichromatic excitation at 589.2\,nm and 569.0\,nm of sodium atoms contained in a buffer-gas-free and uncoated 10-cm-long vapor cell. Directional emissions at $2.34\,\mu$m in the forward direction and $2.21\,\mu$m in both forward and backward directions under different experimental parameters are considered for this study. The measured values of $g^{(2)}(0)$ in all cases are found to exceed unity, 
while remaining significantly below the thermal light limit of 2. 
Cross-correlation measurements reveal that forward- and backward-propagating $2.21\,\mu$m radiations are correlated.
Oscillatory features in $g^{(2)}(\tau)$ are observed over a broad range of excitation powers, and the dependence of the oscillation frequency  on laser power can be attributed to AC Stark shifts, with contributions from hyperfine atomic structure in selected atomic velocity groups even in the presence of Doppler broadening. 
%Our study establishes that the observed mid-infrared emission corresponds to an admixture of lasing with correlated ASE rather than conventional mirrorless lasing, and they highlight the role of coherence buildup and velocity-group selection in shaping photon correlations. 
Our study establishes that the observed mid-infrared emission arises from a phase-matched, continuous-wave cooperative process that combines features of lasing and collective amplified spontaneous emission. 
The results highlight the buildup of long-range dipole coherence and velocity-selective coupling of atomic groups, which together govern the observed photon correlations and forward-backward emission symmetry.
The demonstrated backward emission is of particular interest for applications in laser guidestar generation and mesospheric remote sensing, where understanding the statistical properties of the emitted light is essential for optimizing sodium-based light sources.
\end{abstract}

\maketitle

%\tableofcontents

\section{\label{sec:intro}Introduction}

Coherent and nonlinear optical processes in alkali vapors have a long and rich history, closely linked to the development of laser technology\,\cite{hanna1979,budker2002,pitz17}. 

This progress has enabled the discovery and detailed study of a wide range of phenomena, including Amplified Spontaneous Emission (ASE), Superradiance and Superfluorescence, Stimulated Electronic Raman Scattering (SERS), Four-Wave Mixing (FWM), and others\,\cite{sorokin67,wang69,boyd87,grischkowsky77,Sebbag2019,Londero2009,Zhdanov2011,Andreev}. 
Among these optical processes, mirrorless lasing (ML) in atomic vapors represents a particularly striking example, in which directional coherent emission arises without the use of conventional optical cavities or mirrors.
In this case, the required feedback originates from ASE or distributed feedback within the medium itself, typically supported by population inversion in specific atomic levels.  The term "mirrorless lasing" is sometimes extended to encompass stimulated emission processes that exhibit a distinct pump threshold and spectral narrowing, yet result in a state of relatively low temporal coherence when compared to conventional cavity-based lasers\,\cite{cao03}.

Mirrorless lasing simplifies the design of light sources and makes them attractive for applications in remote sensing, atomic clocks, and precision metrology\,\cite{PhysRevResearch.7.013292}. It is particularly useful for stand-off detection of atmospheric pollutants, trace gases, or environmental parameters\,\cite{zhang2022}. Its high sensitivity, spectral selectivity, and potential for compact, alignment-free setups open pathways for portable and field-deployable sensing technologies\,\cite{kai20,akulshin25}.

Mirrorless lasing in atomic vapors has been realized in various configurations. It was observed for the first time in Cs vapors in Ref.\,\cite{jacobs61} and followed by a number of experiments in alkali vapors leading to pulsed\,\cite{sorokin67,sorokin69} and continuous-wave (CW) lasing\,\cite{sharma81}. 
Degenerate mirrorless lasing was experimentally observed in forward direction in a dilute rubidium vapor under single-linearly polarized CW excitation of the D$_2$ cycling transitions, characterized by orthogonally polarized emission above a $\sim 3$\,mW/cm$^2$ threshold and a sharp magnetic field sensitive resonance\,\cite{Papoyan19}. Non-degenerate ML was observed at 5.2\,$\mu$m in Rb vapors that are stepwise excited by low-power CW resonant light\,\cite{Akulshin14}. 
In Ref.\,\cite{Akulshin18} backward-directed continuous-wave emission at 2.21\,$\mu$m from sodium vapor via two-photon excitation of the 3S$_{1/2} \rightarrow $\,4D$_{5/2}$ transition was demonstrated, achieving a sub 10\,mW threshold. Later, polychromatic ML in hot cesium vapor was observed by pumping the 6S$_{1/2}\rightarrow$ 8P$_{3/2}$ line, producing at least seven infrared emissions and two blue emissions\,\cite{Antypas19}.  This directional blue emission is due to FWM, as there is no population inversion on these transitions, so ASE does not occur. Recent theoretical studies predicted degenerate mirrorless lasing in warm alkali vapors, identifying a Doppler-resilient gain sideband under linearly polarized two-level pumping and outlining conditions for sustained inversion-enhanced gain\,\cite{Ramaswamy23,Ramaswamy25}. Spectral properties of ML, such as temporal coherence and linewidth, can be estimated from the spectral properties of the optical field generated by parametric mixing of ML with the applied pump radiation. Assuming uncorrelated optical fields, the linewidth of ML at 5.23\,$\mu$m in rubidium vapor was estimated to be no more than 1.3\,MHz, though this field has not been directly detected\,\cite{Akulshin14_2}.

Recent theoretical work has demonstrated that coherent, directional light emission can also arise collectively from dense, arrays of subwavelength-spaced dipole-coupled emitters without the use of an optical resonator\,\cite{Ritsch25}. Such systems exhibit a transition from subradiant to superradiant emission accompanied by spectral narrowing, further underscoring the role of dipole–dipole interactions and collective coherence in free-space light generation.

In this broader context, superradiance refers to the cooperative spontaneous emission of phase-synchronized dipoles, producing an intensity that scales as $N^{2}$ and a decay rate enhanced by collective coupling \cite{Dicke,Rehler,Ferioli21}. While originally formulated for pulsed, fully inverted samples, related steady-state regimes can arise under CW driving, where long-range dipole correlations persist and yield directional, phase-matched emission\,\cite{Ferioli,Agarwal}. These cooperative effects provide a natural framework for interpreting correlated bidirectional infrared emission in alkali vapors, bridging the gap between ASE-driven mirrorless lasing and genuine superradiant behavior.

A fundamental quantity for characterizing the statistical and coherence properties of a radiation source is the second-order intensity cross-correlation function  $g^{(2)}_{1,2}(\tau)$\,\cite{loudon}.  It is defined by
\begin{equation}
g^{(2)}_{1,2}(\tau)=\frac{\langle I_1(t)\,I_2(t+\tau)\rangle}{ \langle I_1\rangle \langle I_2\rangle},\label{eq:g2}
\end{equation}
where $I_1(t)$ and $I_2(t)$ denote the registered intensities at two distinct detection channels  (e.g., a Hanbury Brown and Twiss  setup\,\cite{brown56,glauber}).  It reduces to the second-order autocorrelation $g^{(2)}(\tau)$, when $\langle I_1\rangle$ and $ \langle I_2\rangle $ refer to the intensities of the same beam. 
The value $g^{(2)}(0)$  provides insight into the photon statistics and temporal coherence of the light source: (i) for coherent light (e.g., an ideal laser), $g^{(2)}(0) = 1$, indicating Poissonian photon statistics; (ii) for thermal light (e.g., from a blackbody or incandescent bulb), $g^{(2)}(0) = 2$, indicating photon bunching and super-Poissonian statistics; and for single-photon sources, $g^{(2)}(0) < 1$, indicating antibunching, a purely quantum phenomenon\,\cite{loudon}. In Ref.\,\cite{PhysRevA.6.2211}, it was shown that the second-order correlation function $g^{(2)}(\tau)$ can reach values up to 2 for spontaneous decay processes when single-photon detectors are used. 
%\AG{It should be noted that measurements performed with single-photon detectors (start-stop experiments) and conventional photodetectors generally yield different values of $g^{(2)}(\tau)$ because they probe distinct observables. Photon-counting detection resolves discrete arrival events on nanosecond timescales and preserves quantum statistical bunching, whereas photodiode detection integrates the optical intensity over its response bandwidth, effectively averaging over many atomic groups and spatial modes. This temporal coarse-graining suppresses the measured bunching amplitude and tends to drive $g^{(2)}(0)$ toward unity. Consequently, sub-thermal values $1<g^{(2)}(0)<2$ obtained with photodetectors primarily reflect multi-mode averaging rather than the absence of super-Poissonian fluctuations.} It was demonstrated that even with a conventional photodiode (i.e., not a photon-counting device), it is still possible to measure the $g^{(2)}$ function (see e.g.\,\cite{kuusela}).  

It was shown that the intensity correlations of input beams can be partially transferred to generated forward-directed light via FWM in $^{85}$Rb vapor\,\cite{Ihn17}. The 420\,nm collimated blue light inherits the bunching of the input fields, and the measured peak correlation $g^{(2)}(0)$ is significantly reduced due to Doppler broadening in the vapor and finite temporal resolution of the measurement. 

Measurements of $g^{(2)}(\tau)$ revealed evidence of an oscillation, which might have been caused by periodic spiking in the intensity\,\cite{Wiersig2009,wang}. Previous studies revealed complex temporal behavior in alkali vapors: Ref.\,\cite{Akulshin20} reported spiking dynamics in the emission from Rb atoms, whereas Ref.\,\cite{Akulshin21} demonstrated oscillatory behavior of the Pearson correlation coefficient between two signals in Na vapors.
Given the linear relationship between the Pearson coefficient and the second-order correlation function, it is expected that $g^{(2)}(\tau)$ should also exhibit oscillatory behavior. This behavior of the intensity correlations has also been observed in hot Rb vapor, also in the presence of Doppler broadening, indicating that specific atomic velocity groups dominate the observed correlations\,\cite{Almeida25}.

While mirrorless lasing, ASE-driven gain, and directional emission in alkali vapors have been explored in a variety of contexts, quantitative measurements of the second-order coherence function for simultaneously forward- and backward-propagating mid-infrared emission remain comparatively scarce. Here, we characterize the temporal correlations of these counter-propagating beams and identify a persistent oscillatory structure in $g^{(2)}(\tau)$ together with nontrivial cross-correlations. These observations indicate that the gain medium cannot be treated only as an amplified spontaneous emission, but instead also exhibits cooperative coupling between counter-propagating modes under threshold excitation. Thus, analysis of the shape and decay of $g^{(2)}(\tau)$ provides access to fundamental characteristics of the radiation, including coherence time, emission threshold, intensity noise, and the underlying light-matter interaction mechanisms. %\AG{In this work, we investigate correlated forward- and backward-directed mid-infrared emission from sodium vapor under continuous two-color excitation. Particular attention is given to the emergence of directional, phase-matched emission sustained by the continuous drive, representing a steady-state cooperative process rather than purely ASE-driven mirrorless lasing. By comparing the measured second-order correlation functions $g^{(2)}(\tau)$ for the mid-infrared emission with those of conventional laser light, we identify the transition from incoherent amplification to a collective, partially coherent regime. This approach allows us to elucidate the statistical and coherence properties of continuous-wave directional cooperative emission, an analogue of steady-state superradiance in an atomic vapor.}

%In this work, we focus on directional backward and forward emissions in sodium vapor, with particular attention to mirrorless lasing driven by ASE. By comparing the measured $g^{(2)}$ functions of ASE-based mirrorless lasing with those of conventional laser radiation, we aim to elucidate the distinctive statistical and coherence properties of these emission processes.

\begin{figure*}[tb]
\centering
\includegraphics[width=0.6\linewidth]{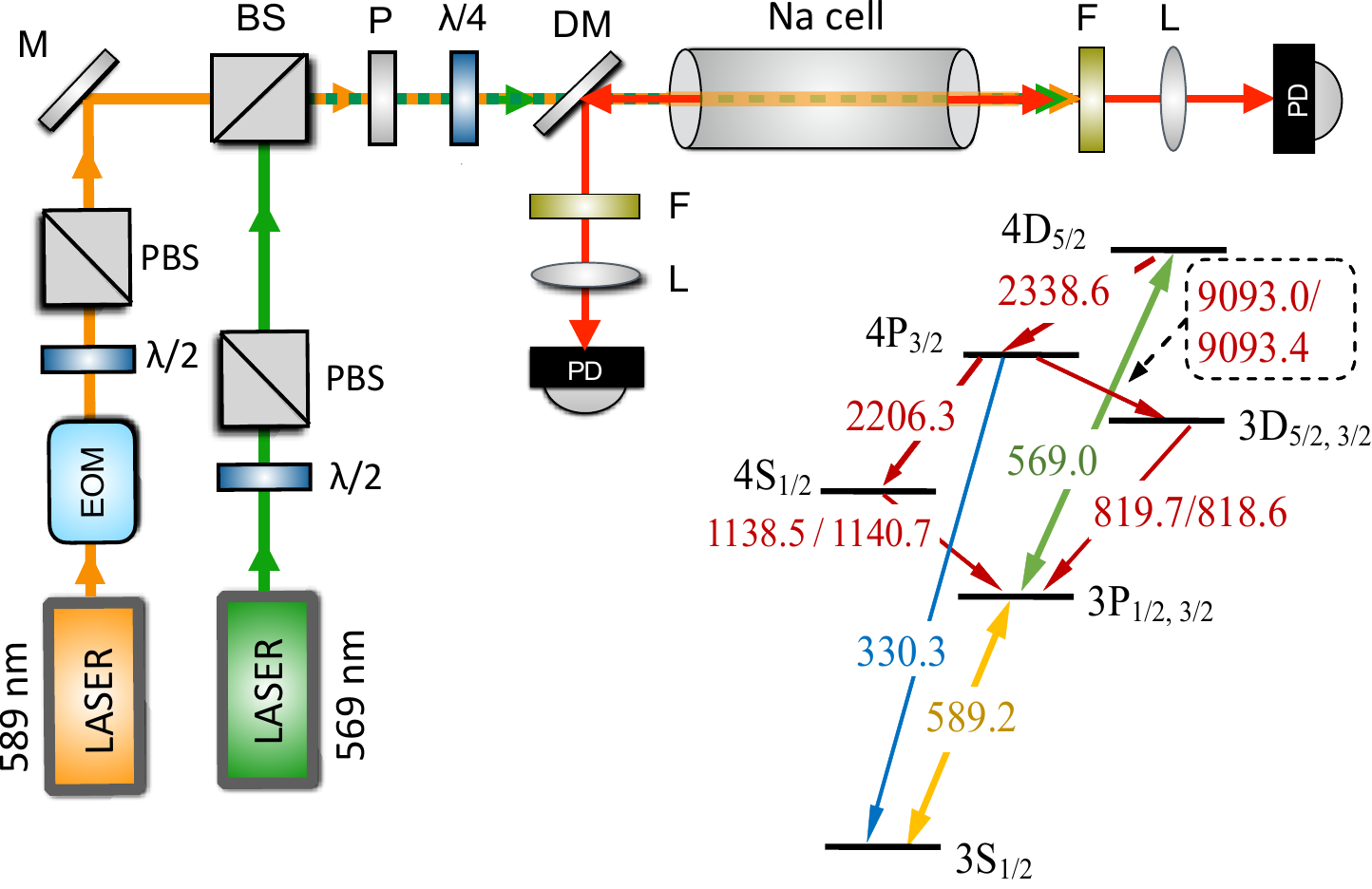}
\caption{\label{fig:setup} Schematic top view of the experimental setup on the left. EOM: Electro-optic modulator driven at 1.713 GHz, PBS: Polarizing beam spliterr, BS: Non-polarizing beam splitter, M: Mirror, P: Polarizer, DM: Dichroic mirror, F: Infrared Filter, L: Lens, PD: Photodetector. %The dashed-line rectangle corresponds to the unit for the measurement of fluorescence. 
Relevant Na energy levels and transitions are presented on the right. %We experimentally observe 2338.6 and 2206.3\,nm radiations. 
}
\end{figure*}

\section{Experimental setup and procedures}

The experimental setup is largely based on the one presented in Ref.\,\cite{akulshin23}. Its simplified version, as well as relevant atomic transitions, are presented in Fig.\,\ref{fig:setup}.

The experiment employs a $10$\,cm long sodium vapor cell without buffer gas. The cell is mounted inside a magnetic shield composed of four-layers of $\mu$-metal to suppress stray magnetic fields; a resistive heater consisting of a pair of twisted wires provides a controlled temperature gradient along the cell while ensuring reduced magnetic field noise. The temperature of the vapor cell is kept at 170$^\circ$C throughout the experiment.  
The wedge-shaped angled  windows of the cell help suppress optical feedback. 

Two laser systems are used for excitation: a solid-state Toptica TA-SHG Pro laser operating at $589.2$\,nm resonant with the $3S_{1/2} \rightarrow 3P_{3/2}(F'=3,2,1)$ transition, and a Coherent 899-21 dye laser at $569.0$\,nm, driving the $3P_{3/2}\rightarrow 4D_{5/2}$ transition.  Efficient ground-state repumping, which lowers the ASE threshold and enhances excitation efficiency\,\cite{Akulshin12}, is achieved by generating spectral sidebands at $1.713\,$GHz via frequency modulation of the $589.2\,$nm laser using an electro-optic modulator (EOM) with a modulation depth of -12\,dBm.
The outputs of the two lasers are combined on a non-polarizing beam splitter to form a bichromatic excitation beam, which is directed into the sodium cell with a minimum spot diameter of approximately 1\,mm. The polarization and intensity of both the $589.2$\,nm and $569.0$\,nm components are controlled using waveplates and polarizers. 

Mid-infrared emission directed along the laser field propagation direction at $2.21\,\mu$m and $2.34\,\mu$m is detected with VIGO LabM-I-10.6 photodetector (PD), and data acquired is registered by a data acquisition (DAQ) system with $500$\,MS/s. We also observe backward emission at $2.21\,\mu$m and at $2.34\,\mu$m. However, the intensity of the $2.34\,\mu$m backward emission is too weak to be reliably detected and analyzed at the same experimental conditions, and therefore it is not further considered in this work. Forward and backward emissions at 2.21\,$\mu$m are simultaneously measured using two PDs, with each radiation recorded on a two-channel DAQ card. For the laser light measurements, we employ a Thorlabs PDA10A photodetector, as the VIGO detector is not sensitive to the laser wavelengths used.

\begin{figure*}[tb]
\centering
\includegraphics[width=0.32\linewidth]{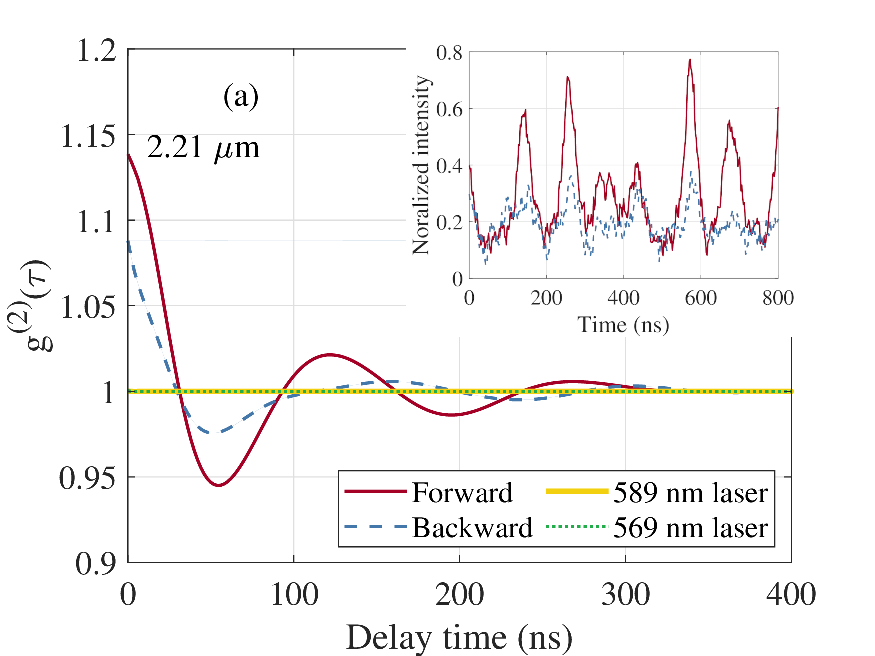}
\includegraphics[width=0.32\linewidth]{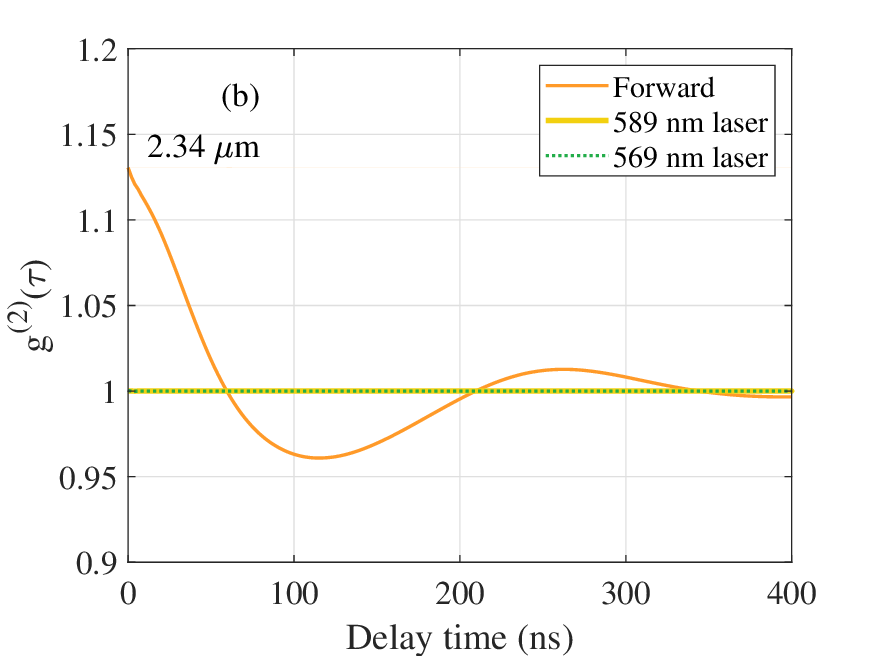}
\includegraphics[width=0.32\linewidth]{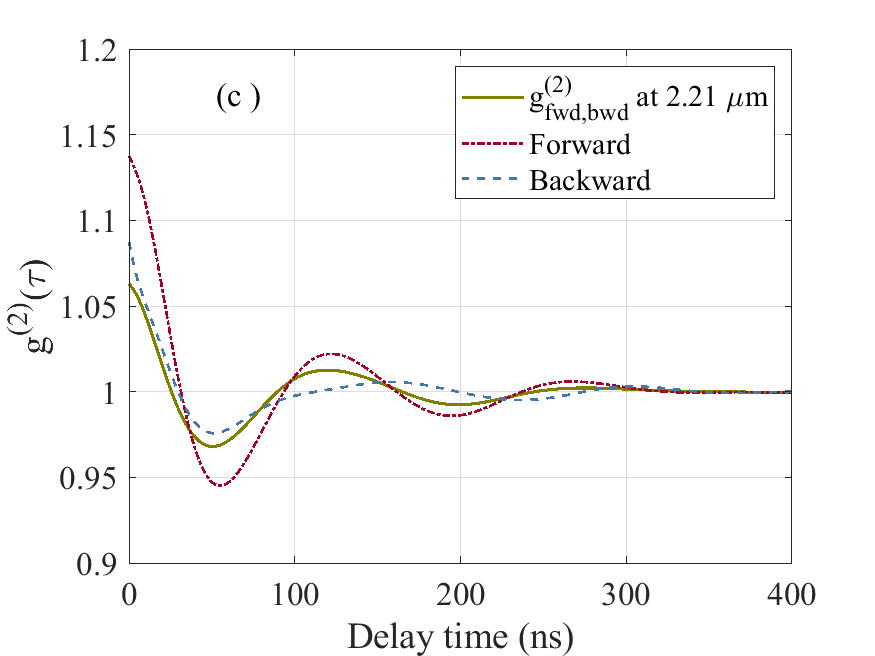}
\caption{\label{fig:g2_tau} Autocorrelation functions vs. delay time $\tau$ for 589.2\,nm laser (solid horizontal line), 569.0\,nm laser (dotted horizontal line), (a) forward (solid) and backward (dashed) emission at $2.21\,\mu$m, and (b) forward emission at $2.34\,\mu$m. (c) Cross-correlation vs. delay time $\tau$ in solid line between forward and backward emitted lights at $2.21\,\mu$m. For reference, $2.21\,\mu$m forward (dash-dotted), and backward (dashed) auto-correlations are shown. The 569.0\,nm laser power is 15\,mW, the 589.2\,nm laser power is 230\,mW. }
\end{figure*}

With 589.2\,nm light at 230\,mW and 569.0\,nm light at 45\,mW, we observe 80\,$\mu$W of the $2.21\,\mu$m collimated light in the forward direction and
45\,$\mu$W in the backward direction. The registered signals exhibit temporal spiking and pronounced oscillations are visible in the traces (see  inset in Fig.\,\ref{fig:g2_tau}a, and Appendix\,\ref{sec:supp_signal}).  

\section{Results and Discussions}

To verify the approach for extracting the second-order correlation functions, we first measure the intensity of CW lasers at 569.0\,nm and 589.2\,nm and calculate the $g^{(2)}(\tau)$ functions, shown in dotted green and solid yellow lines, respectively, in Fig.\,\ref{fig:g2_tau}. The deviation of $g^{(2)}(0)$ from the expected value of 1 is less than $10^{-4}$ for both lasers. All measurements were carried out over a total acquisition time of 500\,ms. The recorded intensity traces were partitioned into 1000 equal segments, and the second-order correlation function $g^{(2)}(\tau)$ was evaluated for each segment and then averaged.

With the same technique (but using the VIGO LabM-I-10.6 detector), we recorded the emission at $2.21\,\mu$m in both forward and backward directions simultaneously, as well as the forward-directed emission at $2.34\,\mu$m. Figure\,\ref{fig:g2_tau}(a) shows the results for the $2.21\,\mu$m emission, with the forward-directed radiation represented by a solid line and the backward-directed radiation by a dashed line. A part of time-dependent raw normalized intensities is shown in Fig.\,\ref{fig:g2_tau}(a) inset, with the forward-directed radiation plotted by solid and the backward-directed radiation by dashed lines, revealing periodic spike structure.
Figure\,\ref{fig:g2_tau}(b) displays the forward-directed emission at $2.34\,\mu$m by a solid line.  
The standard errors are smaller than the line thickness and are invisible in the figures. An example of $g^{(2)}(\tau)$ calculation from a single segment is presented in Appendix\,\ref{sec:supp_signal}.
All data correspond to an incident pump power of $15\,\mathrm{mW}$ at $569.0\,\mathrm{nm}$ and 230\,mW at 589.2\,nm.  As seen in the figure, $g^{(2)}(0) > 1$, indicating photon bunching and demonstrating that the observed "mirrorless lasing" does not correspond to pure lasing,  but rather contains at least an admixture of ASE. Using Eq.\,(3.8.7) from\,\cite{loudon}, the expected value of $g^{(2)}(0)$ for a single-mode signal is expected to be approximately $\sim\,1.78$ for chaotic light, given a coherence time of $\sim\,20\,\text{ns}$, and an integration time of $8\,\text{ns}$. However, our measured value is significantly lower.

\begin{figure}[tb]
\centering
\includegraphics[width=0.48\linewidth]{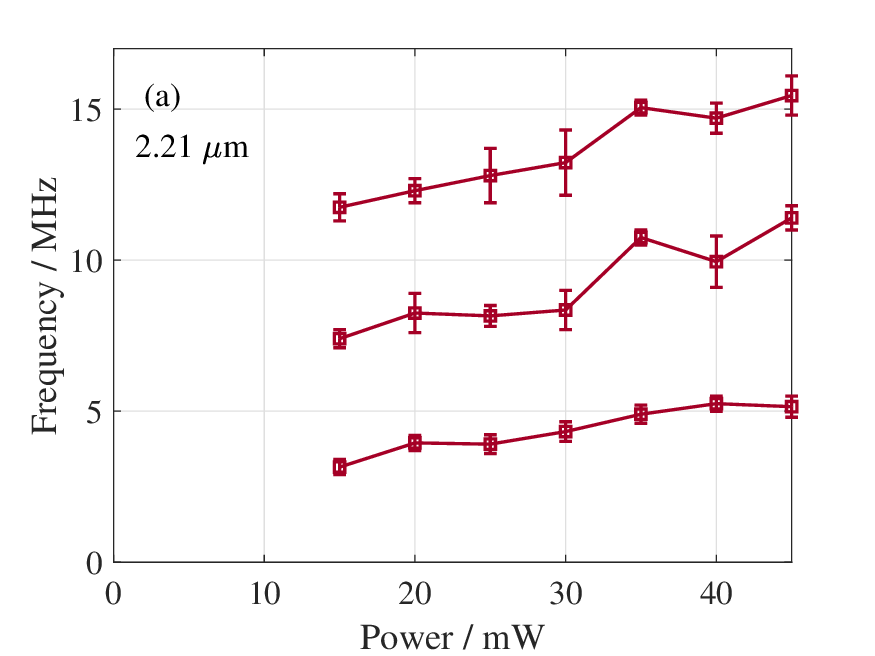}
\includegraphics[width=0.48\linewidth]{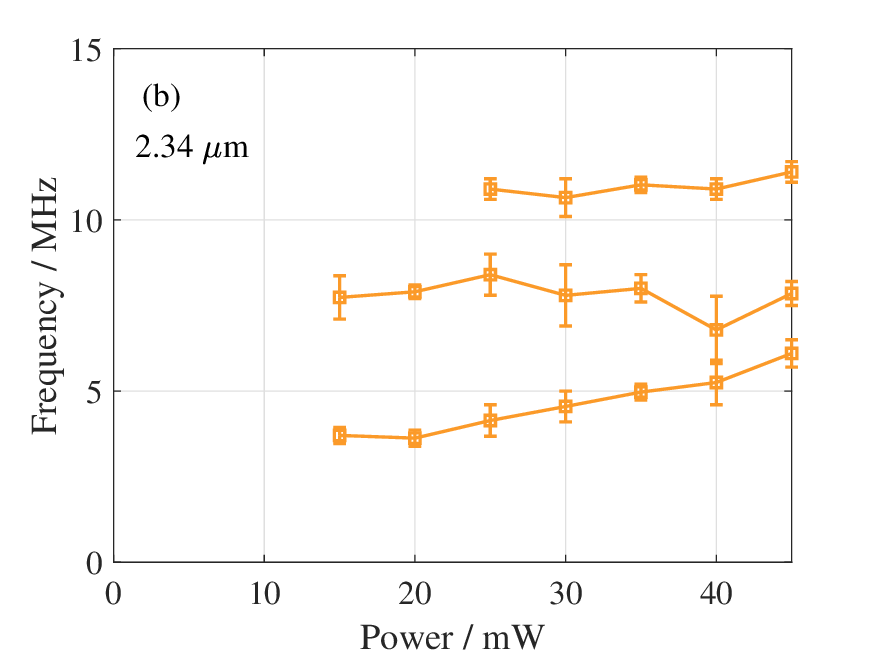}
\caption{\label{fig:frequencies} Dependence of emitted radiation oscillation frequencies on 569.0\,nm laser power for forward emitted (a) 2.21\,$\mu$m and (b) 2.34\,$\mu$m fields. 589.2\,nm laser power is fixed at 230\,mW.}
\end{figure}

On the other hand, sub-thermal values $1<g^{(2)}(0)<2$ can arise from the detection of multiple independent spatio-temporal modes and finite measurement bandwidth (integration time)\,\cite{mika18}. For mutually incoherent chaotic modes, one expects $g^{(2)}(0) = 1 + 1/M$, where $M$ is the number of effective modes\,\cite{Goodman2015}.  

\begin{figure}[h!]
\centering
\includegraphics[width=0.48\linewidth]{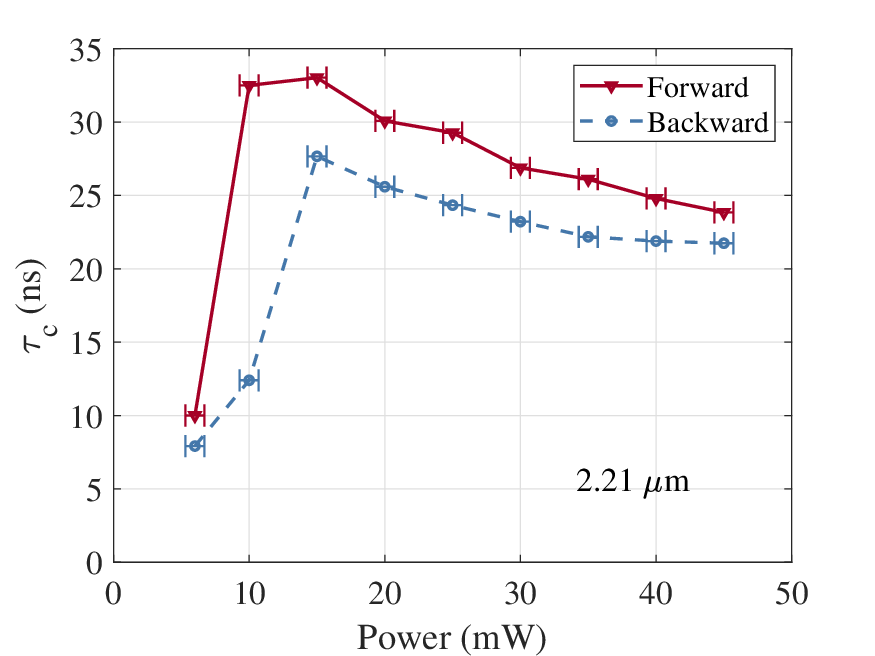}
\includegraphics[width=0.48\linewidth]{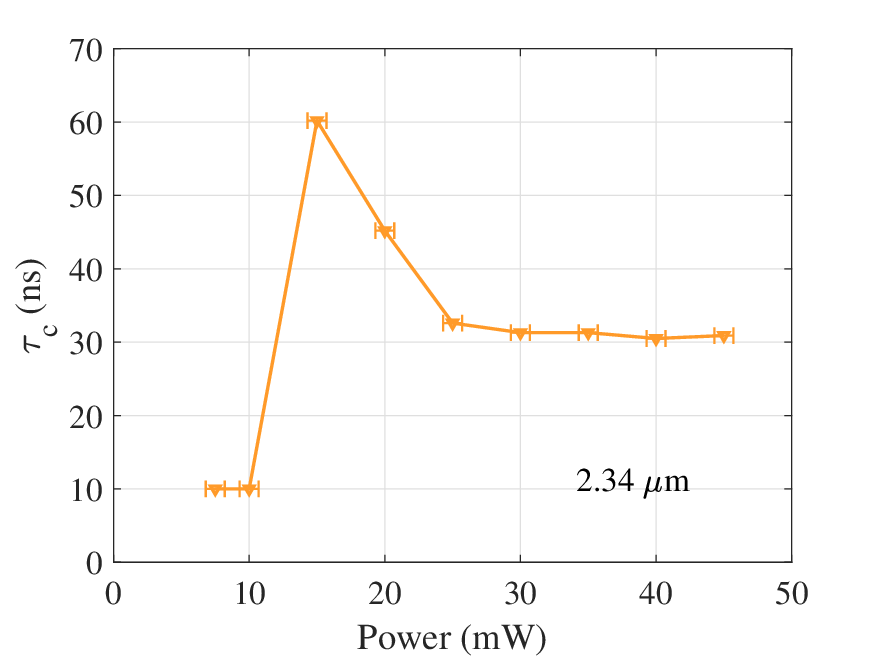}
\includegraphics[width=0.48\linewidth]{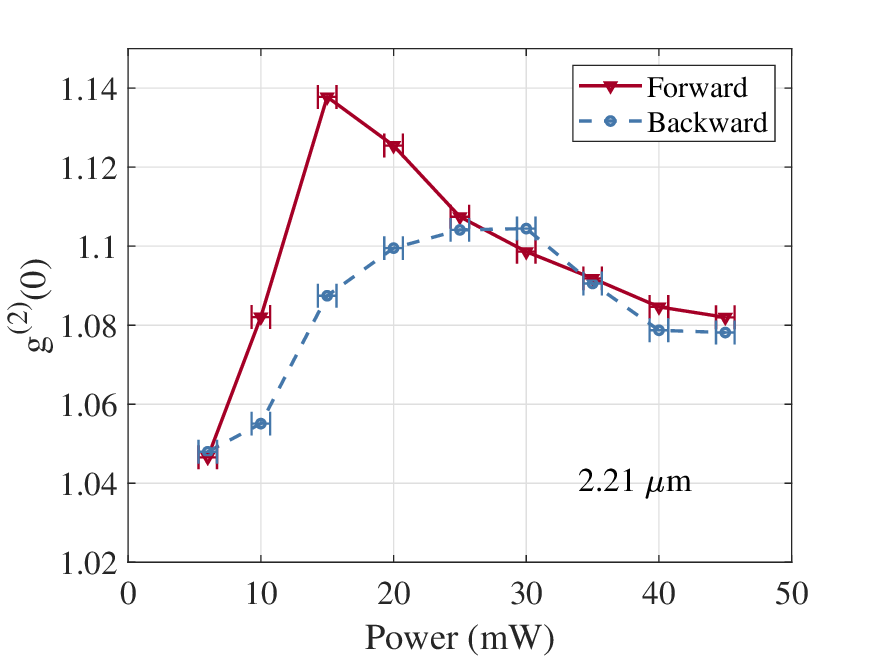}
\includegraphics[width=0.48\linewidth]{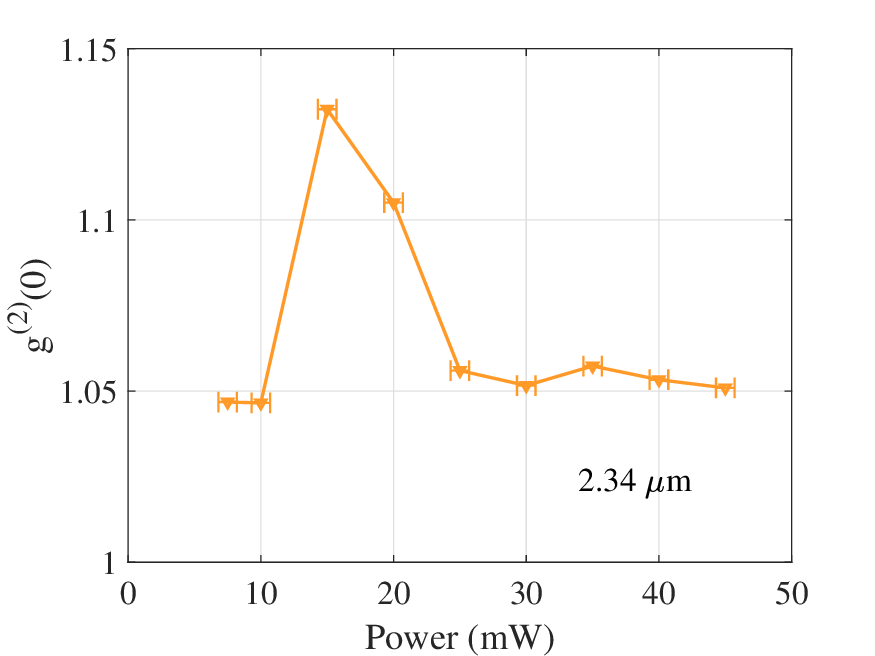}
\includegraphics[width=0.48\linewidth]{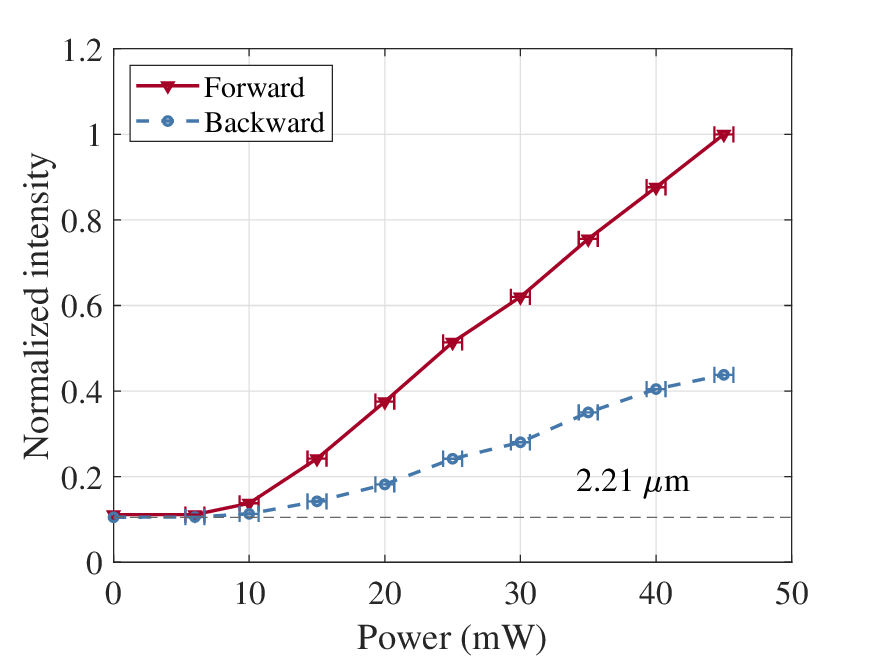}
\includegraphics[width=0.48\linewidth]{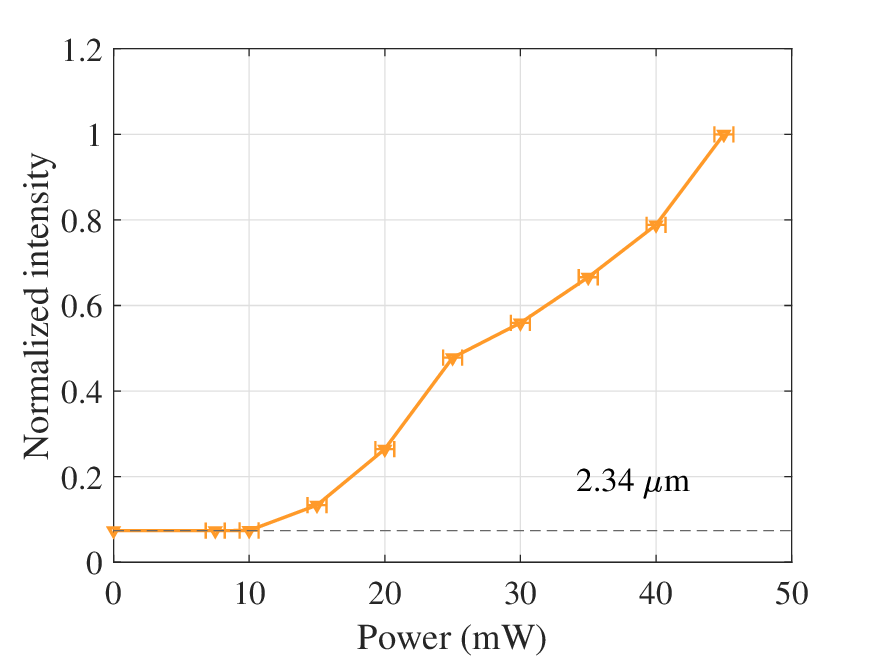}
\caption{\label{fig:221_power} Coherence time (top), $g^{(2)}(0)$ (middle), normalized mean intensity (bottom) of $2.21\,\mu$m of forward (solid) and backward (dashed)  emission (left pane); and $2.34\,\mu$m emitted light (right pane) dependence on 569.0\,nm laser power. 589.2\,nm laser power is 230\,mW. Horizontal error bars indicate the uncertainty of the laser power ($\pm 0.7$\,mW) due to power fluctuations.}
\end{figure}
The calculated cross-correlations between the $2.21\,\mu$m forward- and backward-emitted light is shown in Fig.\,\ref{fig:g2_tau}(c) with a solid curve. 
For reference, the figure includes the auto-correlations of the forward (dash-dotted) and backward (dashed) $2.21\,\mu$m emissions. 
The cross-correlation between the forward- and backward-emitted $2.21\,\mu$m light takes values between one and two. %,  indicating common intensity fluctuation or common seeding process coupling the two directions.
%\AG{It can be understood as a consequence of velocity-selective amplification and optical pumping dynamics within a shared group of atoms. If the beams were generated by statistically independent amplification channels, the cross-correlation would remain near unity within experimental uncertainty.}
This positive cross-correlation between forward and backward channels implies that the two counterpropagating fields are linked through a common macroscopic polarization of the ensemble. This behavior is inconsistent with independent ASE channels, which would yield statistically uncorrelated emission. The correlated infrared radiation observed under continuous two-color excitation, therefore, signals the emergence of a collective, phase-matched radiative process. In this regime, the driving fields imprint a spatially coherent polarization grating across the vapor, leading to constructive interference of the emitted dipole fields along phase-matched directions. The resulting directional enhancement constitutes a CW cooperative emission closely related to superradiance\,\cite{Ferioli}. In contrast to the transient Dicke-type superradiant burst following pulsed inversion, the CW-driven sodium system reaches a steady state in which long-range dipole correlations persist and radiate coherently into both forward and backward modes. %The partially coherent photon statistics, $1<g^{(2)}(0)<2$, and the oscillatory structure of $g^{(2)}(\tau)$ are consistent with steady-state collective emission sustained by continuous driving---a CW analogue of superradiant dynamics.}

In the  dependence on delay time of auto-correlations and cross-correlations, oscillatory behavior is observed. A pronounced oscillation with a period of around 120\,ns, which is equivalent to 8.3\,MHz, is already apparent from the signal structure itself, which is translated to the time-dependence of $g^{(2)}(\tau)$ (see\,Fig.\,\ref{fig:g2_tau}a, solid line). Additional, less obvious components are deduced using the Fourier transform of the signal and  the corresponding $g^{(2)}(\tau)$ function, which yielded consistent spectral features. Oscillation frequencies are presented in Fig.\,\ref{fig:frequencies} as a function of 569.0\,nm laser power, 589.2\,nm laser power is fixed at 230\,mW.   
%\AG{Values $g^{(2)}(\tau) > 1$ correspond to transiently enhanced bunching arising from cooperative population exchange among hyperfine sublevels, while values $g^{(2)}(\tau) < 1$ reflect gain competition and depletion effects in the shared group of atoms. In this picture, velocity-selective amplification and optical pumping periodically favor one propagation direction at the expense of the other, producing the alternating maxima and minima observed in $g^{(2)}(\tau)$.}

Across the range of pump powers investigated, we observe persistent oscillations  around 4--5, 8--11, and 12--16\,MHz, see Fig.\,\ref{fig:frequencies}. 
The systematic upshift of these frequencies with increasing power is consistent with light-induced (AC Stark) shifts of the relevant energy splittings (see Appendix\,\ref{sec:supp_hyperfine}).

Another possible reason for oscillations is the dynamical buildup of emission, arising from the competition between time-coherent mirrorless lasing and ASE processes\,\cite{Wiersig2009,wang}, although in contrast to the results reported in these papers, we do not observe damping of the oscillations far away from the emission threshold. 

Figure\,\ref{fig:221_power} summarizes the dependence of several statistical properties of the forward- and backward-emitted radiation on the 569.0\,nm pump power. The 589.2\,nm laser power was held constant at 230\,mW. The left column corresponds to emission at 2.21\,$\mu$m and the right column to emission at $2.34\,\mu$m. In each column, the top panel displays the coherence time $\tau_c$, the middle panel shows the zero-delay intensity autocorrelation $g^{(2)}(0)$, and the bottom panel reports the normalized mean  intensity of the detected directional emission. For the $2.21\,\mu$m radiation, we plot forward-directed and backward-directed emissions for direct comparison (solid and dotted traces, respectively); the $2.34\,\mu$m emission is shown in the right column. All quantitative fits shown in Fig.\,\ref{fig:221_power} include statistical uncertainties. The $y$-axis uncertainties lie within the symbol size and are therefore not visible, whereas the uncertainty in the laser power is estimated as $\pm 0.7$\,mW, and this is reflected in the horizontal error bars of the plots. 

All measured beams exhibit qualitatively similar behavior as a function of the 569.0\,nm pump power: no directional emission is detected below a pump threshold near $P_{569}\simeq 7$\,mW, while above the threshold, the signals grow and their statistical properties evolve. We note that several different nonlinear optical mechanisms can display threshold-like onsets (e.g. ASE, FWM, and ML). 

The coherence time $\tau_c$ was extracted by fitting the measured correlation function $g^{(2)}(\tau)$ with a Gaussian profile and taking the $1/e$ decay width as $\tau_c$\,\cite{loudon}. The coherence time reaches a maximum at $P_{569}\approx 15$\,mW for all recorded beams, and for larger powers, it approaches approximately 20\,ns for the $2.21\,\mu$m emission and $\sim 30$\,ns for the $2.34\,\mu$m emission. The value of $g^{(2)}(0)$ likewise evolves with 569.0\,nm pump power and asymptotes to a nearly constant value: for the $2.34\,\mu$m emission, this occurs already at $P_{569}\approx 25$\,mW, whereas for the $2.21\,\mu$m emission, the asymptotic behavior is reached at higher pump powers, $\sim 40$\,mW. 
For the $2.21\,\mu$m emission, $g^{(2)}(0)$ rises from about 1.04 at threshold to a peak value of 1.14 before settling to 1.08 at higher powers, while for the $2.34\,\mu$m emission, it follows a similar trend, peaking at 1.13 and stabilizing near 1.05. Such small deviations above unity are consistent with nearly coherent (laser-like) emission, but with a weak excess of super-Poissonian fluctuations that likely arise from residual amplified spontaneous emission or parametric contributions.

The tendency of both $\tau_c$ and $g^{(2)}(0)$ to level off at high 569.0\,nm pump power is consistent with the system reaching a steady-state regime in which the dominant gain and loss processes balance; this behaviour is expected for both chaotic source [$g^{(2)}(0) = 2$] and a laser [$g^{(2)}(0) = 1$].

The mean detected intensity of the directional emission exhibits a clear threshold near $P_{569} \approx 10$ mW, see Fig.\ref{fig:221_power}, bottom row. Below this value, the signal remains constant at the detector offset level (indicated by the thin black dashed line). Above the threshold, the intensity increases linearly with pump power, indicating the onset of gain-dominated directional emission.

\section{Conclusion}

In summary, we have investigated the statistical properties of sodium emission under bichromatic excitation by analyzing the second-order intensity correlation function $g^{(2)}(\tau)$. 
The directional emissions at $2.21$ and $2.34\,\mu$m exhibit $g^{(2)}(0)$ values between one and two, together with strong forward--backward correlations and persistent oscillations, pointing to a phase-matched, steady-state cooperative emission regime. These observations are consistent with CW directional superradiance\,\cite{Ferioli}, in which synchronized dipoles radiate coherently into both counterpropagating modes.
Such behavior naturally accommodates the measured subthermal statistics and the presence of multiple effective modes, as partial mode averaging reduces the observed bunching amplitude without destroying collective coherence. 

Throughout the explored range of pump powers, $g^{(2)}(\tau)$ of the emissions exhibited persistent oscillations at characteristic frequencies near 4--5, 8--11, and 12--16\,MHz, with a systematic upshift of these components at higher excitation power. It is consistent with AC Stark shifts of the underlying hyperfine sublevels participating in the cooperative polarization, a mechanism analogous to multi-mode superradiant beating.

%Measurements of directional emissions at $2.21\,\mu$m and $2.34\,\mu$m in both forward and backward directions reveal values of $g^{(2)}(0)$ that consistently exceed unity, indicating bunching behavior characteristic of an admixture of ASE on top of lasing. The fact that $g^{(2)}(0)$ remains significantly below the thermal light limit of 2 can be explained by partial coherence build-up in the medium, the presence of multiple effective modes, and detection-time limitations. Cross-correlation measurements further demonstrate that forward- and backward-propagating $2.21\,\mu$m radiations are correlated. 
%The oscillatory features observed in $g^{(2)}(\tau)$ persist over a broad range of excitation powers and can be attributed to hyperfine structure of atomic states, and AC Stark shifts, with contributions from specific atomic velocity groups even in the presence of Doppler broadening. 

This behavior can be placed in a broader context: in photon Bose--Einstein condensation the mode populations follow a Bose--Einstein distribution, and the growth of the macroscopic ground-state occupation is reflected in the second-order correlation function $g^{(2)}(\tau)$, which tracks the crossover from thermal to coherent behavior. Near threshold the condensate may still exhibit bunching with $g^{(2)}(0) > 1$ due to reservoir-induced number fluctuations, while in a well-developed condensate the ideal expectation is $g^{(2)}(0) \approx 1$, signifying second-order coherence\,\cite{schmitt14}. Although directional scattering and photon condensation arise from distinct physical mechanisms--non-equilibrium amplification versus equilibrium condensation--both phenomena can be characterized by their photon correlations. In this sense, $g^{(2)}(\tau)$ offers a unifying diagnostic for identifying thermal, amplified, and condensed regimes of light, even though the statistical origins of coherence differ.

In addition, the backward-emitted light offers promising opportunities for laser guide star generation and mesospheric remote sensing, and our $g^{(2)}$ analysis provides valuable insight into the coherence and mode structure of this emission, paving the way for the optimization of sodium-based light sources for atmospheric and astronomical applications. Looking ahead, extending these studies to more complex excitation geometries and structured light fields could provide further insight into the interplay between coherence, polarization, and spatial mode structure. In particular, measuring quantum correlations in polychromatic radiation generated by twisted light carrying orbital angular momentum (OAM) represents a promising direction for future research.

\section*{Acknowledgements}
Authors acknowledge helpful discussions with M.\,Chekhova, M.\, Fleischhauer, J.\,Marino, H.\,Ritsch, S.\,Malinovskaya, A.\,Nomerotski, H.\,Karapetyan, A.\,Wickenbrock, H.\,Bekker, M.\,Khanbekyan, E.\,Klinger. 

\section*{Funding}

This work was supported by the Higher Education and Science Committee of the Republic of Armenia (Research projects N24-2PTS-1C003 and N24-2PTS-1C004), through the QuantERA II Program, with funding received via the EU H2020 research and innovation program under Grant No. 101017733 and associated support from the German VDI under Grant No. 13N16931 (V-MAG), project HEU-RIA-MUQUABIS  Grant No. 101070546.

\bibliography{bib_file}% Produces the bibliography via BibTeX.

\appendix

\section{\label{sec:supp_signal}Temporal structure and auto- and cross-correlation functions}

\begin{figure}[tbh]
\centering
\includegraphics[width=0.48\linewidth]{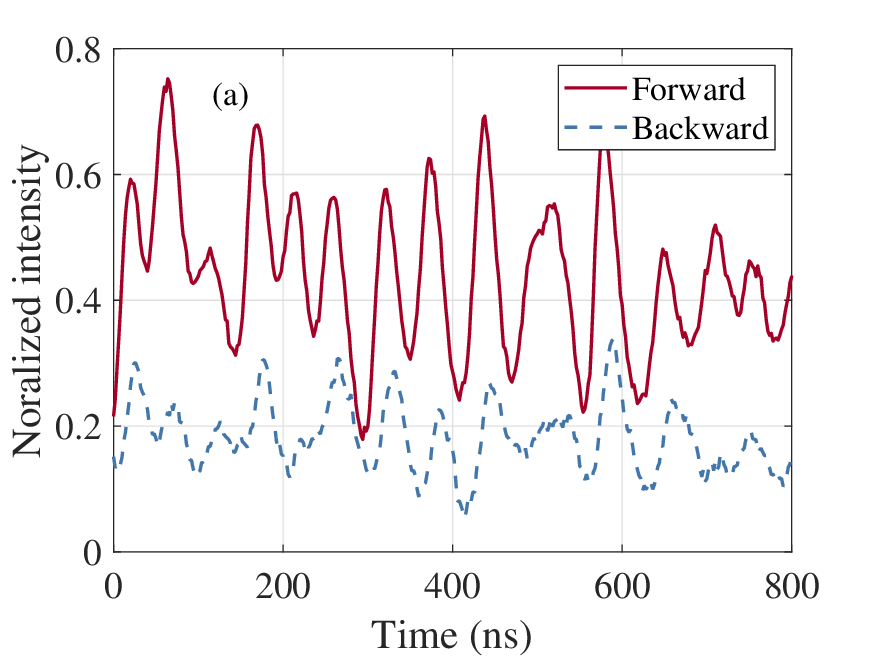}
\includegraphics[width=0.49\linewidth]{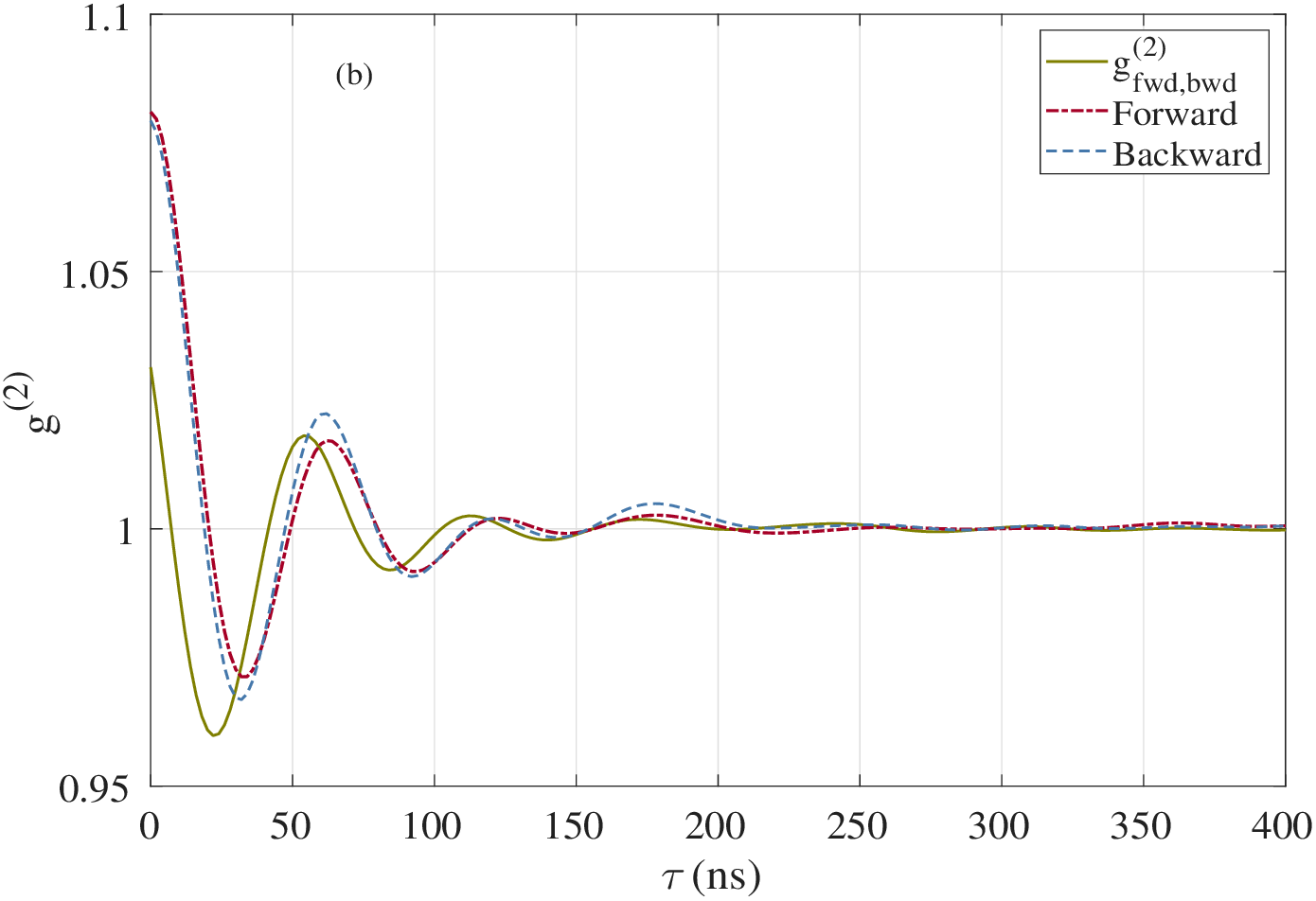}
\caption{\label{fig:S_signal} (a) Time dependence of the detected intensities at 2.21\,$\mu$m in the forward (solid) and backward (dashed) directions, recorded with laser powers of 230\,mW at 589.2\,nm and 45\,mW at 569.0 nm. (b) Corresponding second-order auto-correlation functions $g^{(2)}(\tau)$ calculated from the same data: forward emission (dash-dotted line), backward emission (dashed line), and forward-backward cross-correlation (solid).%, shown for both positive and negative delay times. The slight shift of the cross-correlation maximum toward negative delays ($\tau \simeq -8$\,ns) is attributed to instrumental timing asymmetries between detection channels.
}
\end{figure}

In order to illustrate the temporal dynamics of the detected mid-infrared emission, we present the raw intensity traces of the 2.21\,$\mu$m radiation in both forward and backward directions under excitation with 230\,mW at 589.2\,nm and 45\,mW at 569.0\,nm. The time-dependent intensities are shown in Fig.\,\ref{fig:S_signal}(a), with the forward-directed radiation plotted by solid and the backward-directed radiation by dashed lines. The radiations clearly exhibit spike-like structures in both the forward and backward directions, which appear largely in phase, indicating a strong correlation between the two emissions.  The mean of the signals of the forward and backward emissions is in the order of $\sim 10^3$, and variance is in the order of $\sim 10^5$.

The corresponding second-order correlation functions, calculated directly from these data, are presented in Fig.\,\ref{fig:S_signal}(b). Here, the dash-dotted line corresponds to the forward emission, the dashed line to the backward emission, and the solid curve to the cross-correlation between the two. %\AG{Unlike the symmetric auto-correlation functions, the cross-correlation peak exhibits a small temporal shift toward negative delays, approximately -8\,ns at a pump power of 45\,mW at 569.0\,nm and -4\,ns at lower pump powers, although we have ensured we use identical cables of identical lengths.}
%\AG{This offset most likely originates from instrumental effects such as unequal group delays in detector response, small differences in photodetector bandwidth or rise time, or a finite channel skew between the two input channels of the DAQ system. Since the optical paths for forward and backward emissions are nearly identical and the propagation distances are short compared with the nanosecond scale, genuine physical asymmetry in emission timing can be excluded. The slight reduction of the apparent delay at lower pump power may reflect power-dependent variations in detector response time or in the effective triggering synchronization between channels. Therefore, we attribute the observed time shift in the cross-correlation to an instrumental timing offset rather than a physical delay in the emission process.}
These results provide a representative example of the raw data underlying the correlation analysis and demonstrate the presence of correlated fluctuations between forward- and backward-propagating emissions at 2.21\,$\mu$m.

\section{\label{sec:supp_hyperfine}Hyperfine splittings of relevant states}

As discussed in the main text, the temporal structure of the registered emission signals reveals several characteristic frequency components. Across the investigated 569.0\,nm laser power range, we observe persistent oscillations in  $g^{(2)}(\tau)$ around 4--5, 8--11, and 12--16\,MHz. For comparison, the hyperfine splittings of the $4D_{3/2,5/2}$, $4P_{1/2,3/2}$, and $4S_{1/2}$ states were estimated using data from Refs.\,\cite{Arimondo77,Biraben78,Burghardt78}, and the corresponding energy separations $\Delta F$ are summarized in Table\,\ref{tab:I}. The highlighted row indicates the transitions whose hyperfine differences most closely match the observed oscillation frequencies, suggesting that these arise from beatings between the respective sublevels. Discrepancies can be attributed to AC Stark shifts, finite accuracy of hyperfine constants, and experimental factors such as detection bandwidth, residual magnetic fields, and multimode coupling.

\begin{table}[h]
\centering
\begin{tabular}{c c c c c}
\hline
$\Delta F$ & 1--0 & 2--1 & 3--2 & 4--3 \\
\hline
$4D_{3/2}$ & 0.23  & 0.46   & 0.69   & $\times$ \\
$4D_{5/2}$ & $\times$ & 0.056  & 0.084  & 0.11 \\
$4P_{1/2}$ & $\times$ & 54     & $\times$ & $\times$ \\
\rowcolor{lightgray} $4P_{3/2}$ & 5.1   & 11.2   & 19.3   & $\times$ \\
$4S_{1/2}$ & $\times$ & 404    & $\times$ & $\times$ \\
\hline
\end{tabular}
\caption{Estimated hyperfine splittings $\Delta F$ for the $4D_{3/2,5/2}$, $4P_{1/2,3/2}$, and $4S_{1/2}$ states of sodium. The highlighted row indicates transitions whose hyperfine separations correspond most closely to the oscillation frequencies observed in $g^{(2)}(\tau)$.}\label{tab:I}
\end{table}

\end{document}